\documentclass[]{jpsj2}

\def\runauthor{J. \textsc{Otsuki}, H. \textsc{Kusunose} and Y. \textsc{Kuramoto}}

\title{%
Dynamics of the Singlet-Triplet System Coupled with Conduction Spins\\
-- Application to Pr Skutterudites --
}

\author{%
Junya \textsc{Otsuki}\thanks{E-mail address: otsuki@cmpt.phys.tohoku.ac.jp},
Hiroaki \textsc{Kusunose} and Yoshio \textsc{Kuramoto}
}

\inst{%
Department of Physics, Tohoku University, Sendai 980-8578
}

\recdate{\today}

\abst{%
Dynamics of the singlet-triplet crystalline electric field (CEF) system at finite temperatures is discussed by use of the non-crossing approximation. 
Even though the Kondo temperature is smaller than excitation energy to the CEF triplet, the Kondo effect appears at temperatures higher than the CEF splitting, and accordingly only quasi-elastic peak is found in the magnetic spectra. 
On the other hand, at lower temperatures the CEF splitting suppresses the Kondo effect and inelastic peak develops.
The broad quasi-elastic neutron scattering spectra observed in PrFe$_4$P$_{12}$ at temperatures higher than the quadrupole order correspond to the parameter range where the CEF splittings are unimportant. 
}

\kword{%
Kondo effect, dynamical susceptibility, inelastic neutron scattering, non-crossing approximation (NCA), filled skutterudite, PrFe$_4$P$_{12}$, PrOs$_4$Sb$_{12}$
}

\begin{document}
\maketitle
%
%
\section{Introduction}
Heavy fermion states have been observed in some Pr skutterudites such as PrOs$_4$Sb$_{12}$ and PrFe$_4$P$_{12}$. 
PrOs$_4$Sb$_{12}$ is the first heavy fermion superconductor discovered in Pr compounds\cite{Bauer} and, in addition, has quadrupole ordered phase under high magnetic field\cite{Aoki_OsSb}.
PrFe$_4$P$_{12}$ exhibits antiferro-quadrupole order at low temperatures\cite{Aoki_FeP}. 
In the inelastic neutron scattering experiment of PrFe$_4$P$_{12}$, any CEF excitation peak cannot be detected at temperatures higher than the phase transition, while inelastic peak develops at lower temperatures\cite{Iwasa_Fe}. 
On the other hand, the CEF excitations are clearly seen in PrOs$_4$Sb$_{12}$\cite{Maple, kuwahara, gore}. 
Taking account of the fact that only PrFe$_4$P$_{12}$ exhibits the Kondo effect in Pr skutterudite series\cite{Sato}, 
we consider that the heavy mass in PrFe$_4$P$_{12}$ is caused by the Kondo effect, while another mechanism may be relevant to PrOs$_4$Sb$_{12}$. 

CEF level structures are essential to the properties of $4f$ electrons. 
The CEF singlet ground state and low lying excited triplet states make clear the origin of the high field ordered phase in PrOs$_4$Sb$_{12}$\cite{Kohgi,shiina-aoki}. 
The ordered phase in PrFe$_4$P$_{12}$ is also ascribed to the quadrupole moment of the CEF singlet-triplet system\cite{Kiss-Fazekas}. 
Difference of the actual wave functions of the CEF triplet states should create the diversity of physical properties in Pr skutterudite. 

According to the band calculation, main conduction band in Pr skutterudites is formed by the $2p$ orbitals of pnictogens with $a_u$ symmetry, which does not have the orbital degrees of freedom\cite{Harima}. 
Considering exchange interaction between $4f^2$ states and $a_u$ orbital, we have revealed the condition for the occurrence of the Kondo effect \cite{otsuki}. Then the fact that the Kondo effect appears only in PrFe$_4$P$_{12}$ is reasonably understood by the difference between the triplet wave functions. 

When the Kondo effect occurs in the singlet-triplet system, the Kondo effect competes with the CEF singlet.  
It is expected that the competition affects the physical quantities, especially their temperature dependence. 
The Kondo effect in the CEF singlet ground state has been studied theoretically by several authors in relation to URu$_2$Si$_2$\cite{kuramoto,shimizu,yotsuhashi}. However, its dynamics such as dynamical susceptibility has not been derived in spite of its significance. 

In this paper, we derive dynamics of the CEF singlet-triplet system, and clarify temperature dependence of dynamical quantities. The competition between the CEF singlet and the Kondo effect is the main issue. 
Dynamics is derived by the non-crossing approximation (NCA) in \S 2. 
Numerical results by the NCA and by the numerical renormalization group are given in \S 3. Then we discuss application to the real materials. 
We summarize the results in the last section. 
Details of the NCA equations for an actual computation are given in Appendices A and B.

%
%
\section{Application of the NCA to the Singlet-Triplet System}
The NCA has been developed as a solver of the Anderson model\cite{nca1,bickers}.  The NCA enables us to compute dynamical quantities at finite temperature as well as static ones. 
An equivalent approximation has been applied to the Coqblin-Schrieffer model with CEF splittings, and temperature dependence of dynamical susceptibility has been discussed\cite{maekawa}. 
We shall newly apply the NCA to the localized $4f^2$ system with exchange interactions. 

%
%
\subsection{Singlet-triplet Kondo model}
We have derived the effective exchange interaction in the singlet-triplet system coupled with the $a_u$ conduction spin $\mib{s}_c$ \cite{otsuki}. The interaction is written in terms of operator $\mib{X}^{\rm t}$ and $\mib{X}^{\rm s}$, or pseudo-spins $\mib{S}_1 =(\mib{X}^{\rm t}+\mib{X}^{\rm s})/2$ and $\mib{S}_2 =(\mib{X}^{\rm t}-\mib{X}^{\rm s})/2$ as follows:
\begin{align}
	H_{\text{s-t}} &= 
	\epsilon_{\rm t} P_{\rm t}+
	\epsilon_{\rm s} P_{\rm s}+
	\left( 
	 I_{\rm t} \mib{X}^{\rm t} + I_{\rm s} \mib{X}^{\rm s}
	  \right)\cdot \mib{s}_c \nonumber \\
	&=  
	\Delta_{\rm CEF} \mib{S}_1\cdot \mib{S}_2 +
	(J_1\mib{S}_1 + J_2\mib{S}_2)\cdot \mib{s}_c,
\label{eq:st-kondo}
\end{align}
where $P_{\rm s}=-\mib{S}_1 \cdot\mib{S}_2 +1/4$ and $P_{\rm t}=\mib{S}_1 \cdot\mib{S}_2 +3/4$ are the projection operator onto singlet and triplet states, respectively. $J_{1, 2}=I_{\rm t} \pm I_{\rm s}$ and $\Delta_{\rm CEF}=\epsilon_{\rm t}-\epsilon_{\rm s}$ is the CEF splitting. We have chosen the origin of energy so that $3\epsilon_{\rm t} +\epsilon_{\rm s}=0$.
In the second order perturbation on hybridization, the coupling constant $J_1$ becomes ferromagnetic and $J_2$ antiferromagnetic, provided the triplet wave functions mainly consist of $|\Gamma_4\rangle$ in the point group $O_h$. On the other hand, both exchange interactions can be negligible if the triplet is almost composed of $|\Gamma_5\rangle$ in $O_h$.
\begin{figure}[thb]
	\begin{center}
	\includegraphics[width=5cm]{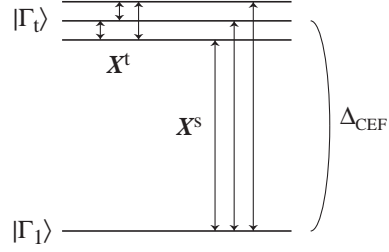}
	\end{center}
	\caption{Operators acting on the singlet-triplet system.}
	\label{fig:x_operator}
\end{figure}

The operator $\mib{X}^{\rm t}$ act on the triplet states, and $\mib{X}^{\rm s}$ connect the singlet and the triplet as shown in Fig. \ref{fig:x_operator}.
In applying the NCA to the singlet-triplet Kondo model, it is convenient to work in the form of $\mib{X}^{\rm t}$ and $\mib{X}^{\rm s}$ because singlet and triplet are properly distinguished from each other. 

\subsection{Integral equations for resolvents and effective interactions}
We derive dynamics of the singlet-triplet system adopting the NCA.
In applying the NCA to the exchange interactions, we introduce a fictitious $4f^1$ intermediate state with negligible population\cite{bickers}. 
We could alternatively introduce a fictitious $4f^3$ interdediate state instead of the $4f^1$ state. 

\begin{figure}[tb]
	\begin{center}
	\includegraphics[width=7cm]{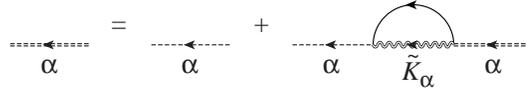}
	\end{center}
	\caption{Diagrammatical representation of the equation for the resolvent and self-energy in the NCA. Solid, dashed and wavy line denote conduction electron, resolvent and interaction, respectively. Single and double line represent bare and renormalized one, respectively.}
	\label{fig:dyson_resolv}
\end{figure}

We introduce resolvent $R_{\rm s}(z)$ for the singlet state and $R_{\rm t}(z)$ for the triplet.
The effects of the interactions are taken account of as the self-energy $\Sigma_{\alpha}(z)$.
Each resolvent $R_{\alpha}(z)$ is given by
\begin{align}
	R_{\alpha}(z) = [z-\epsilon_{\alpha} - \Sigma_{\alpha}(z)]^{-1},
\end{align}
where $\alpha$ denotes configurations of $4f^2$ states.
The NCA determines $\Sigma_{\alpha}(z)$ in a self-consistent fashion. 
Figure \ref{fig:dyson_resolv} shows diagrammatical representation of the equation for the resolvent and self-energy in the NCA. 

\begin{figure}[tb]
	\begin{center}
	\includegraphics[width=15cm]{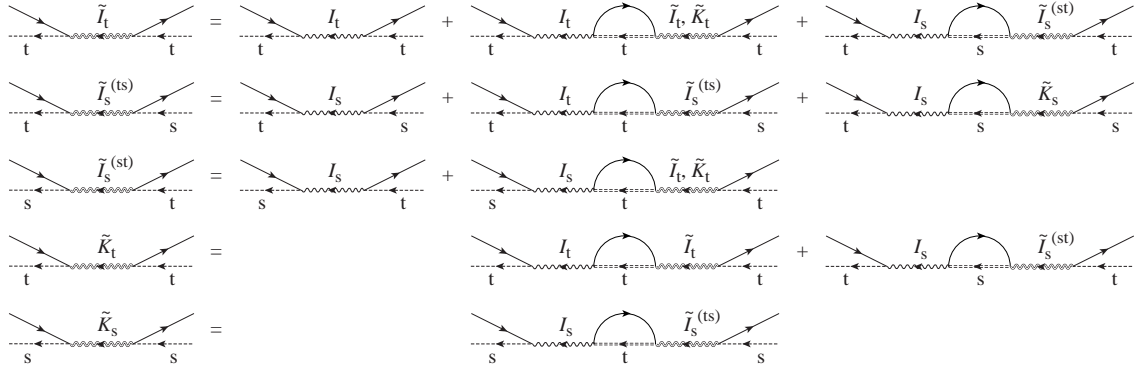}
	\end{center}
	\caption{Diagrammatical representations of the equations for the effective interactions in the NCA.}
	\label{fig:dyson_interaction}
\end{figure}

We obtain the renormalized exchanges $\tilde{I}_{\rm t}(z)$ and $\tilde{I}_{\rm s}(z)$, which are modified from the bare ones $I_{\rm t}$ and $I_{\rm s}$. 
In addition, an effective potential $\tilde{K}_{\rm t}(z)$ for the triplet and $\tilde{K}_{\rm s}(z)$ for the singlet are generated by higher-order exchange scatterings. 
The self-energy is given in terms of the effective potentials by the NCA integral equation:
\begin{align}
	\Sigma_{\alpha}(z) = -2 \int d\epsilon \rho_c(\epsilon) [1-f(\epsilon)]
	 \tilde{K}_{\alpha} (z-\epsilon).
\label{eq:self}
\end{align}
In order to derive equations for the renormalized interactions $\tilde{I}_{\alpha}(z)$ and $\tilde{K}_{\alpha}(z)$, we divide the operator $\mib{X}^{\rm s}$ into two parts:
\begin{align}
	\mib{X}^{\rm s} = P_{\rm t}\mib{X}^{\rm s}P_{\rm s} + P_{\rm s}\mib{X}^{\rm s}P_{\rm t}.
\end{align}
The first term operates on the singlet state, and second one to the triplet states.
Correspondingly, we define effective interactions $\tilde{I}_{\rm s}^{\rm (ts)}(z)$ and $\tilde{I}_{\rm s}^{\rm (st)}(z)$ for each part. 
The effective interactions are determined by equations illustrated in Fig. \ref{fig:dyson_interaction}.
The simultaneous equations for the effective interactions are given in the matrix form by
\begin{align}
	\left(
	\begin{array}{c}
		\tilde{I}_{\rm t} \\ \tilde{I}_{\rm s}^{\text{(ts)}} \\ \tilde{I}_{\rm s}^{\text{(st)}} \\
		\tilde{K}_{\text{t}} \\ \tilde{K}_{\text{s}}
	\end{array} \right) = \left(
	\begin{array}{c}
		I_{\rm t} \\ I_{\rm s} \\ I_{\rm s} \\ 0 \\ 0
	\end{array} \right) - \left(
	\begin{array}{ccccc}
		I_{\rm t} \Pi_{\text{t}}/2 & 0 & I_{\rm s} \Pi_{\text{s}}/2 & I_{\rm t} \Pi_{\text{t}} & 0 \\
		0 & I_{\rm t} \Pi_{\text{t}} & 0 & 0 & I_{\rm s} \Pi_{\text{s}} \\
		I_{\rm s} \Pi_{\text{t}} & 0 & 0 & I_{\rm s} \Pi_{\text{t}} & 0 \\
		I_{\rm t} \Pi_{\text{t}}/2 & 0 & I_{\rm s} \Pi_{\text{s}}/4 & 0 & 0 \\
		0 & I_{\rm s} \Pi_{\text{t}} 3/4 & 0 & 0 & 0
	\end{array}
	\right) \left(
	\begin{array}{c}
		\tilde{I}_{\rm t} \\ \tilde{I}_{\rm s}^{\text{(ts)}} \\ \tilde{I}_{\rm s}^{\text{(st)}} \\
		\tilde{K}_{\text{t}} \\ \tilde{K}_{\text{s}}
	\end{array} \right),
\label{eq:dyson_interaction}
\end{align}
where auxiliary quantity $\Pi_{\alpha}(z)$ are introduced by
\begin{align}
	\Pi_{\alpha}(z) = \int d \epsilon \rho_c (\epsilon) f(\epsilon) R_{\alpha}(z+\epsilon),
\label{eq:pi}
\end{align}
which appears as a bubble diagram in Fig. \ref{fig:dyson_interaction}.
In deriving eq. (\ref{eq:dyson_interaction}), we have used the following properties: 
\begin{align}
	&\sum_{\alpha_1 \sigma_1}
	(\mib{X}^{\rm t}_{\alpha \alpha_1} \cdot \mib{\sigma}_{\sigma_1 \sigma'})
	(\mib{X}^{\rm t}_{\alpha_1 \alpha'} \cdot \mib{\sigma}_{\sigma \sigma_1})
	= (2 P_{\text{t}} \sigma^0 + \mib{X}^{\rm t} \cdot \mib{\sigma})_{\alpha \alpha', \sigma \sigma'}, \nonumber \\
	&\sum_{\alpha_1 \sigma_1}
	(\mib{X}^{\rm s}_{\alpha \alpha_1} \cdot \mib{\sigma}_{\sigma_1 \sigma'})
	(\mib{X}^{\rm s}_{\alpha_1 \alpha'} \cdot \mib{\sigma}_{\sigma \sigma_1})
	= (P_{\text{t}} \sigma^0 + 3 P_{\text{s}} \sigma^0
	 + \mib{X}^{\rm t} \cdot \mib{\sigma})_{\alpha \alpha', \sigma \sigma'}, \nonumber \\
	&\sum_{\alpha_1 \sigma_1}
	(\mib{X}^{\rm t}_{\alpha \alpha_1} \cdot \mib{\sigma}_{\sigma_1 \sigma'})
	(\mib{X}^{\rm s}_{\alpha_1 \alpha'} \cdot \mib{\sigma}_{\sigma \sigma_1})
	= 2P_{\text{t}}\mib{X}^{\rm s}_{\alpha \alpha'}P_{\text{s}} \cdot \mib{\sigma}_{\sigma \sigma'}, \nonumber \\
	&\sum_{\alpha_1 \sigma_1}
	(\mib{X}^{\rm s}_{\alpha \alpha_1} \cdot \mib{\sigma}_{\sigma_1 \sigma'})
	(\mib{X}^{\rm t}_{\alpha_1 \alpha'} \cdot \mib{\sigma}_{\sigma \sigma_1})
	= 2P_{\text{s}}\mib{X}^{\rm s}_{\alpha \alpha'}P_{\text{t}} \cdot\mib{\sigma}_{\sigma \sigma'},
\end{align}
where $\sigma^0$ denotes a unit operator for the conduction spin. 
Solving eq. (\ref{eq:dyson_interaction}) for the renormalized interactions, we obtain
\begin{align}
	\tilde{I}_{\rm t} &= \frac{4(2I_{\rm t} - I_{\rm s}^2 \Pi_{\text{s}}) }
	 {(2-I_{\rm t} \Pi_{\text{t}})(4 + 4 I_{\rm t} \Pi_{\text{t}} - 3I_{\rm s}^2 \Pi_{\text{t}}\Pi_{\text{s}})}
	 \equiv \frac{c_1}{ab}, \nonumber \\
	\tilde{I}_{\rm s} &\equiv \tilde{I}_{\rm s}^{\text{(ts)}} =  \tilde{I}_{\rm s}^{\text{(st)}}
	 = \frac{ 4I_{\rm s} }{(4 + 4 I_{\rm t} \Pi_{\text{t}} - 3I_{\rm s}^2 \Pi_{\text{t}}\Pi_{\text{s}})}
	 \equiv \frac{c_2}{b}, \nonumber \\
	\tilde{K}_{\text{t}} &= -\frac{4I_{\rm t}^2 \Pi_{\text{t}} + 2I_{\rm s}^2 \Pi_{\text{s}}
	 - 3 I_{\rm t} I_{\rm s}^2 \Pi_{\text{t}} \Pi_{\text{s}} }
	{(2-I_{\rm t} \Pi_{\text{t}})(4 + 4 I_{\rm t} \Pi_{\text{t}} - 3I_{\rm s}^2 \Pi_{\text{t}}\Pi_{\text{s}})}
	 \equiv -\frac{c_3}{ab}, \nonumber \\
	\tilde{K}_{\text{s}} &= -\frac{3 I_{\rm s}^2\Pi_{\text{t}}}
	{4 + 4 I_{\rm t} \Pi_{\text{t}} - 3I_{\rm s}^2 \Pi_{\text{t}}\Pi_{\text{s}}}
	 \equiv -\frac{c_4}{b},
\label{eq:renorm_interaction}
\end{align}
where the simplifying notation $a=2-I_{\rm t} \Pi_{\text{t}}$, $b=4 + 4 I_{\rm t} \Pi_{\text{t}} - 3I_{\rm s}^2 \Pi_{\text{t}}\Pi_{\text{s}}$, etc. will be utilized in Appendix A. 
The effective interactions $\tilde{I}_{\alpha}(z)$ are required only by the vertex corrections for the dynamical magnetic susceptibility.

%
%
\subsection{$T$-matrix}
In the NCA for the exchange interactions, the impurity $T$-matrix is computed as follows:
\begin{align}
	T(i \epsilon_n) = - \frac{1}{Z_f} \int_C \frac{dz}{2\pi i} e^{-\beta z}
	 \sum_{\alpha} \tilde{K}_{\alpha}(z) R_{\alpha}(z+i\epsilon_n),
	\label{eq:t-matrix}
\end{align}
where $\epsilon_n=(2n+1)\pi T$ is the Matsubara frequency of fermions and the contour $C$ encircles all singularities of the integrand counter-clockwise. 
We utilize the spectral intensities
\begin{align}
	\eta_{\alpha}(\omega) &= - \frac{1}{\pi} \text{Im} R_{\alpha}(\omega + i\delta), \\
	\xi_{\alpha}(\omega) &= Z_f^{-1} e^{-\beta \omega} \eta_{\alpha}(\omega),
	\label{eq:tilde_xi}
\end{align}
where $\delta$ is positive infinitesimal.
We further define $\eta_{\alpha}^{\rm (K)}(\omega)$ and $\xi_{\alpha}^{\rm (K)}(\omega)$ for $\tilde{K}_{\alpha}(z)$, and $\eta_{\alpha}^{\rm (I)}(\omega)$ and $\xi_{\alpha}^{\rm (I)}(\omega)$ for $\tilde{I}_{\alpha}(z)$ in a similar fashion. 
The spectral function $\xi_{\alpha}(\omega)$ should be actually computed by another set of equations to avoid difficulty of the Boltzmann factor at low temperatures. We shall describe the details in Appendix A.
Performing analytic continuation $i\epsilon_n \rightarrow \omega +i\delta$ to real frequencies in eq. (\ref{eq:t-matrix}), we obtain
\begin{align}
	-\frac{1}{\pi} \text{Im}T(\omega + i \delta)
	= \sum_{\alpha} \int d\epsilon [\xi_{\alpha}^{(K)} (\epsilon) \eta_{\alpha}(\epsilon + \omega)
	 + \eta_{\alpha}^{(K)} (\epsilon) \xi_{\alpha}(\epsilon + \omega)].
\end{align}
Electrical conductivity is derived from $\text{Im}T(\omega)$ as described later.

%
%
\subsection{Dynamical magnetic susceptibility}
The dynamical magnetic susceptibility is given by
\begin{align}
	&\chi(i \nu_m) 
	= (g_J \mu_{\text{B}})^2 \sum_{\alpha \alpha' \beta \beta'}
	 \langle \alpha' | J_z | \beta \rangle \langle \beta' | J_z | \alpha \rangle 
	 \chi (\alpha \alpha', \beta \beta'; i \nu_m), \\
	&\chi (\alpha \alpha', \beta \beta'; i \nu_m)
	 = -\frac{1}{Z_f} \int_C \frac{dz}{2 \pi i} e^{-\beta z}
	 \Lambda (\alpha \alpha', \beta \beta'; z, z+i \nu_m),
	\label{eq:chi_nu}
\end{align}
where $\nu_m=2m\pi T$ is the boson Matsubara frequency, and $\Lambda(\alpha \alpha', \beta \beta'; z, z+i \nu_m)$ is the vertex function to be explained below. 
The operator $J_z$ has non-zero matrix elements for
\begin{align}
	\langle \text{t}\pm | J_z | \text{t}\pm \rangle = \pm M_{\rm t},\ 
	\langle \text{s} | J_z | \text{t0} \rangle = M_{\rm s}.
\label{eq:Jz}
\end{align}
Although $M_{\rm t}$ and $M_{\rm s}$ depend on the triplet wave functions in the point group $T_h$ \cite{shiina-aoki}, 
we take these matrix elements as unity in the following numerical calculations. 
It is straightforward to reinstate the actual values for particular wave functions. 

\begin{figure}[tb]
	\begin{center}
	\includegraphics[width=6cm]{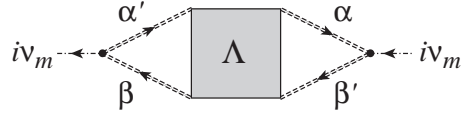}
	\end{center}
	\caption{Diagrammatical representation of the dynamical magnetic susceptibility.}
	\label{fig:vertex}
\end{figure}

\begin{figure}[tb]
	\begin{center}
	\includegraphics[width=15cm]{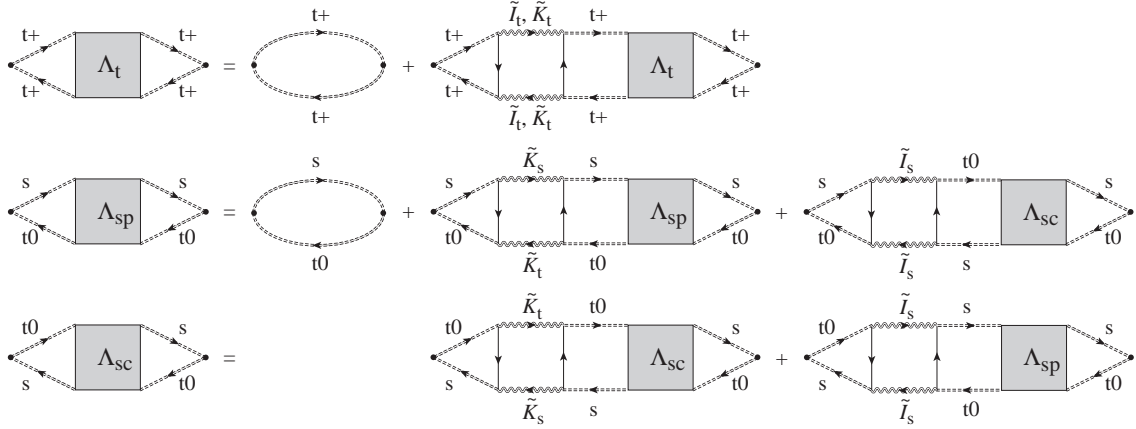}
	\end{center}
	\caption{Diagrammatical representation of the equations for the dynamical magnetic susceptibility.}
	\label{fig:vertex_eq}
\end{figure}

The vertex part $\Lambda(\alpha \alpha', \beta \beta'; z, z+i \nu_m)$ is diagrammatically represented in Fig. \ref{fig:vertex}.
It contains infinite pairs of conduction electrons and holes through the Bethe-Salpeter-type equations in the NCA. 
There are six combinations $(\alpha \alpha', \beta \beta')$ that contribute to the dynamical susceptibility. A combination is symbolically represented by $\lambda$ hereafter. 
Only three kinds of vertices are independent (see Table \ref{tab:vertex}).
We note that $\langle \text{t}+ | J_z | \text{t}+ \rangle$ and $\langle \text{t}- | J_z | \text{t}- \rangle$ are completely decoupled owing to cancellation between the processes through the state $|\text{t}0 \rangle$.
\begin{table}[tb]
	\caption{Combinations $(\alpha \alpha', \beta \beta')$ and relevant quantities for the dynamical magnetic susceptibility.}
	\begin{center}
	\begin{tabular}{|c|c|c|}
		\hline
		$\alpha \alpha', \beta \beta'$ &
		 $\langle \alpha' | J_z | \beta \rangle \langle \beta' | J_z | \alpha \rangle$ &
		 $\Lambda(\alpha \alpha', \beta \beta'; z, z+i \nu_m)$ \\
		\hline
		t+t+, t+t+ & $|\langle \text{t}+ | J_z | \text{t}+ \rangle |^2$ &
		 $\Lambda_{\rm t} (z, z+i \nu_m)$ \\
		t$-$t$-$, t$-$t$-$ & $|\langle \text{t}- | J_z | \text{t}- \rangle |^2$ &
		 $\Lambda_{\rm t} (z, z+i \nu_m)$ \\
		s s, t0 t0 & $|\langle \text{t}0 | J_z | \text{s} \rangle |^2$ &
		 $\Lambda_{\rm sp} (z, z+i \nu_m)$ \\
		t0 t0, s s & $|\langle \text{t}0 | J_z | \text{s} \rangle |^2$ &
		 $\Lambda_{\rm sp'} (z, z+i \nu_m) = \Lambda_{\rm sp} (z+i \nu_m, z)$ \\
		s t0, s t0 & $\langle \text{t}0 | J_z | \text{s} \rangle ^2$ &
		 $\Lambda_{\rm sc} (z, z+i \nu_m)$ \\
		t0 s, t0 s & $\langle \text{s} | J_z | \text{t}0 \rangle ^2$ &
		 $\Lambda_{\rm sc'} (z, z+i \nu_m) = \Lambda_{\rm sc} (z+i \nu_m, z)$ \\
		\hline
	\end{tabular}
	\end{center}
	\label{tab:vertex}
\end{table}
Figure \ref{fig:vertex_eq} shows diagrammatical representation of the equations for the vertices.
The equations are analytically given by
\begin{align}
	\Lambda_{\rm t} (z,z') = R_{\rm t}(z) R_{\rm t}(z')
	+ R_{\rm t}(z) R_{\rm t}(z') 2\int d\epsilon \rho_c (\epsilon) [1-f(\epsilon)]
	 \int d\epsilon' \rho_c (\epsilon') f(\epsilon') \nonumber \\
	 \times \left[ \frac14 \tilde{I}_{\rm t} (z-\epsilon) \tilde{I}_{\rm t} (z'-\epsilon)
	  +\tilde{K}_{\rm t} (z-\epsilon) \tilde{K}_{\rm t} (z'-\epsilon) \right]
	 \Lambda_{\rm t} (z-\epsilon+\epsilon',z'-\epsilon+\epsilon'),
	 \label{eq:ver_t}
\end{align}
\begin{align}
	\Lambda_{\rm sp} (z,z') = R_{\rm s}(z) R_{\rm t}(z')
	+ R_{\rm s}(z) R_{\rm t}(z') 2\int d\epsilon \rho_c (\epsilon) [1-f(\epsilon)]
	 \int d\epsilon' \rho_c (\epsilon') f(\epsilon') \nonumber \\
	 \times [ \tilde{K}_{\rm s} (z-\epsilon) \tilde{K}_{\rm t} (z'-\epsilon)
	 \Lambda_{\rm sp} (z-\epsilon+\epsilon',z'-\epsilon+\epsilon') \nonumber \\
	+ \frac14 \tilde{I}_{\rm s} (z-\epsilon) \tilde{I}_{\rm s} (z'-\epsilon)
	 \Lambda_{\rm sc} (z-\epsilon+\epsilon',z'-\epsilon+\epsilon') ],
	 \label{eq:ver_sp}
\end{align}
\begin{align}
	\Lambda_{\rm sc} (z,z') =
	R_{\rm t}(z) R_{\rm s}(z') 2\int d\epsilon \rho_c (\epsilon) [1-f(\epsilon)]
	 \int d\epsilon' \rho_c (\epsilon') f(\epsilon') \nonumber \\
	 \times [ \tilde{K}_{\rm t} (z-\epsilon) \tilde{K}_{\rm s} (z'-\epsilon)
	 \Lambda_{\rm sc} (z-\epsilon+\epsilon',z'-\epsilon+\epsilon') \nonumber \\
	+ \frac14 \tilde{I}_{\rm s} (z-\epsilon) \tilde{I}_{\rm s} (z'-\epsilon)
	 \Lambda_{\rm sp} (z-\epsilon+\epsilon',z'-\epsilon+\epsilon') ].
	 \label{eq:ver_sc}
\end{align}
The other vertices $\Lambda_{\rm sp'} (z,z')$ and $\Lambda_{\rm sc'} (z,z')$ satisfy the equations with replacement $R_{\rm t} \leftrightarrow R_{\rm s}$, $\tilde{K}_{\rm t} \leftrightarrow \tilde{K}_{\rm s}$, $\Lambda_{\rm sp} \rightarrow \Lambda_{\rm sp'}$, and $\Lambda_{\rm sc} \rightarrow \Lambda_{\rm sc'}$ in eqs. (\ref{eq:ver_sp}) and (\ref{eq:ver_sc}), respectively. 

After analytic continuation $i\nu_m \rightarrow \omega +i\delta$ to the real axis in eq. (\ref{eq:chi_nu}), we obtain
\begin{align}
	&\text{Im} \chi_{\lambda}(\omega +i\delta)
	 = (1- e^{-\beta \omega})
	\frac{1}{Z_f} \int_{-\infty}^{\infty} \frac{d\epsilon}{2\pi} e^{-\beta \epsilon} \nonumber \\
	&\times  \text{Re} [\Lambda_{\lambda}(\epsilon -i\delta, \epsilon +\omega +i\delta)
	  -\Lambda_{\lambda}(\epsilon +i\delta, \epsilon +\omega +i\delta)].
\label{eq:Im_chi}
\end{align}
The static susceptibility $\chi_{\lambda}(0)$ is given by
\begin{align}
	\chi_{\lambda}(0) =
	 \frac{1}{Z_f} \int_{-\infty}^{\infty} \frac{d\epsilon}{\pi} e^{-\beta \epsilon}
	 \text{Im} \Lambda_{\lambda}(\epsilon +i\delta, \epsilon +i\delta).
\label{eq:chi0}
\end{align}
In numerical calculations, it is difficult to perform the integration including the Boltzmann factor in eqs. (\ref{eq:Im_chi}) and (\ref{eq:chi0}).
We avoid the difficulty through transforming eqs. (\ref{eq:ver_t}), (\ref{eq:ver_sp}) and (\ref{eq:ver_sc}) into equivalent and convenient forms (see Appendix B).

%
%
\section{Numerical Results for Resistivity and Susceptibility}
The singlet-triplet Kondo model has two characteristic energy scales, i.e., the Kondo temperature $T_{\rm K}$ and renormalized CEF splitting $\tilde{\Delta}_{\rm CEF}$.
Provided the CEF singlet has lower energy than the CEF triplet, the Kondo effect and the CEF singlet compete with each other.
We shall examine how the competition affects the temperature dependence of dynamical quantities. 
We first show the phase diagram of the ground state by the numerical renormalization group (NRG), and next dynamical quantities obtained by the NCA. 

%
%
\subsection{Numerical renormalization group analysis}
The singlet-triplet Kondo model has four fixed points, i.e., CEF singlet, CEF triplet, doublet and quartet. 
They have residual entropies $\ln1 =0$, $\ln3$, $\ln2$ and $\ln4$, respectively. 
Some explicit results for temperature dependence of the entropy by the NRG have been presented in ref. \citen{otsuki}.
\begin{figure}[tb]
	\begin{center}
	\includegraphics[width=7cm]{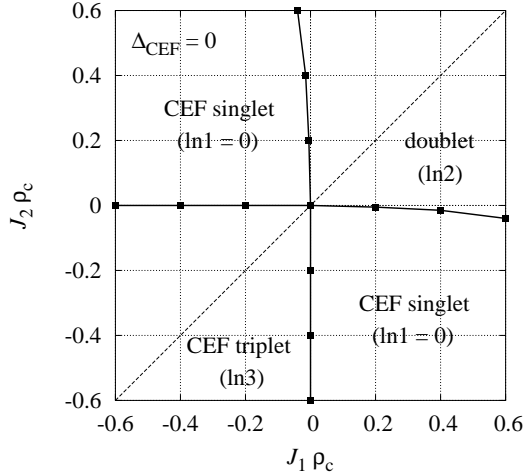}
	\end{center}
	\caption{$J_1$-$J_2$ phase diagram of the ground state with $\Delta_{\rm CEF}=0$.}
	\label{fig:phase_diagram}
\end{figure}
Figure \ref{fig:phase_diagram} shows the $J_1$-$J_2$ phase diagram of the ground state of the singlet-triplet Kondo model with the bare CEF splitting $\Delta_{\rm CEF}=0$. 
Note the symmetry against interchange of $J_1$ and $J_2$. 
The effective CEF splitting becomes finite by renormalization effect. 
In the second order perturbation theory, self-energies of the triplet and singlet states are written as
\begin{align}
	\Sigma^{(2)}_{\rm t}(z) &\sim -\frac{1}{2} D\rho_c^2 (2I_{\rm t}^2 + I_{\rm s}^2), \nonumber \\
	\Sigma^{(2)}_{\rm s}(z) &\sim -\frac{3}{2} D\rho_c^2 I_{\rm s}^2, 
\end{align}
at $T=0$ and $|z| \ll D$. 
Then we obtain the renormalized CEF splitting $\Delta^{(2)}_{\rm CEF}$ in the second order perturbation theory as
\begin{align}
	\Delta^{(2)}_{\rm CEF} = -D \rho_c^2 (I_{\rm s}^2 -I_{\rm t}^2)
	 = -D \rho_c^2 J_1 J_2.
	 \label{eq:energy_shift}
\end{align}
It is now clear that with $J_1J_2<0$, the second order exchange stabilize the CEF singlet. 
On the other hand, higher order exchanges cause the Kondo effect, which makes the CEF splitting obscure. 
The competition between the Kondo effect and the CEF effect depends on their characteristic energy scales. 
We find that renormalization of the CEF splitting is larger than the Kondo temperature in the calculated parameter range. 

In computing dynamical quantities, we take $J_2=0$ to minimize renormalization of the CEF splitting. 
Even if $J_2$ equal zero, pseudo-spin $\mib{S}_2$ would interact with conduction electrons indirectly through another pseudo-spin $\mib{S}_1$. 
The singlet-triplet Kondo model with the additional condition $\Delta_{\rm CEF}=0$ is reduced to the Kondo model where the pseudo-spin $\mib{S}_2$ is decoupled. 

\subsection{Electrical resistivity}
We discuss how the CEF singlet influences the Kondo effect, comparing with the behavior in the CEF triplet ground state. 
We take a rectangular model with the band width $2D$ for conduction electrons
\begin{align}
	\rho_c(\epsilon) = \theta(D-|\epsilon|)/2D,
\end{align}
where $\theta(\epsilon)$ is a step function. 
We tentatively assume $D=10^4$K and take Kelvin as the unit of energy.

\begin{figure}[thb]
	\begin{center}
	\includegraphics[width=7cm]{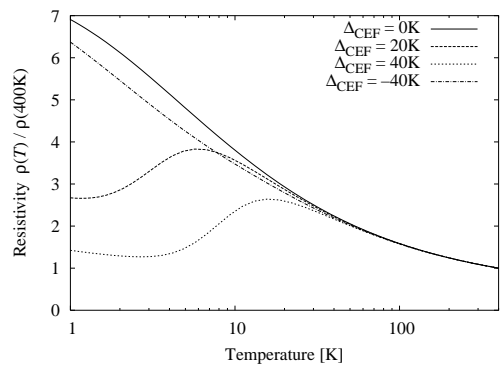}
	\end{center}
	\caption{The electrical resistivity as a function of temperature for several values of $\Delta_{\rm CEF}$.}
	\label{fig:resistivity}
\end{figure}

We obtain the electrical conductibity $\sigma(T)$ by
\begin{align}
	\sigma(T)= A \int d\epsilon \left( -\frac{\partial f(\epsilon)}{\partial \epsilon} \right)
	\frac{1}{|\text{Im}T(\epsilon)|},
\end{align}
where $A$ is a constant and $f(\epsilon)$ is the Fermi distribution function. 
Figure \ref{fig:resistivity} shows temperature dependence of the electrical resistivity $\rho(T)=1/\sigma(T)$ with $J_1\rho_c =0.2$. 
We take $\Delta_{\rm CEF}$ ranging from $-40$K (triplet) to 40K (singlet). 
In the case of the triplet ground state, resistivity monotonously increases as temperature decreases, corresponding to development of the Kondo resonance peak. The CEF splitting only causes a slight reduction of the slope in comparison with the quartet case $\Delta_{\rm CEF}=0$. 
On the other hand, the CEF singlet leads to suppression of the enhancement at temperatures below about the third of $\Delta_{\rm CEF}$. 
The suppression is due to a pseudo-gap of about $2\Delta_{\rm CEF}$ in $\text{Im}T(\omega)$ around Fermi level. 
Note that the peak temperature of $\rho(T)$ is substantially smaller than $\Delta_{\rm CEF}$. 

\subsection{Dynamical magnetic susceptibility}
\begin{figure}[thb]
	\begin{center}
	\includegraphics[width=7cm]{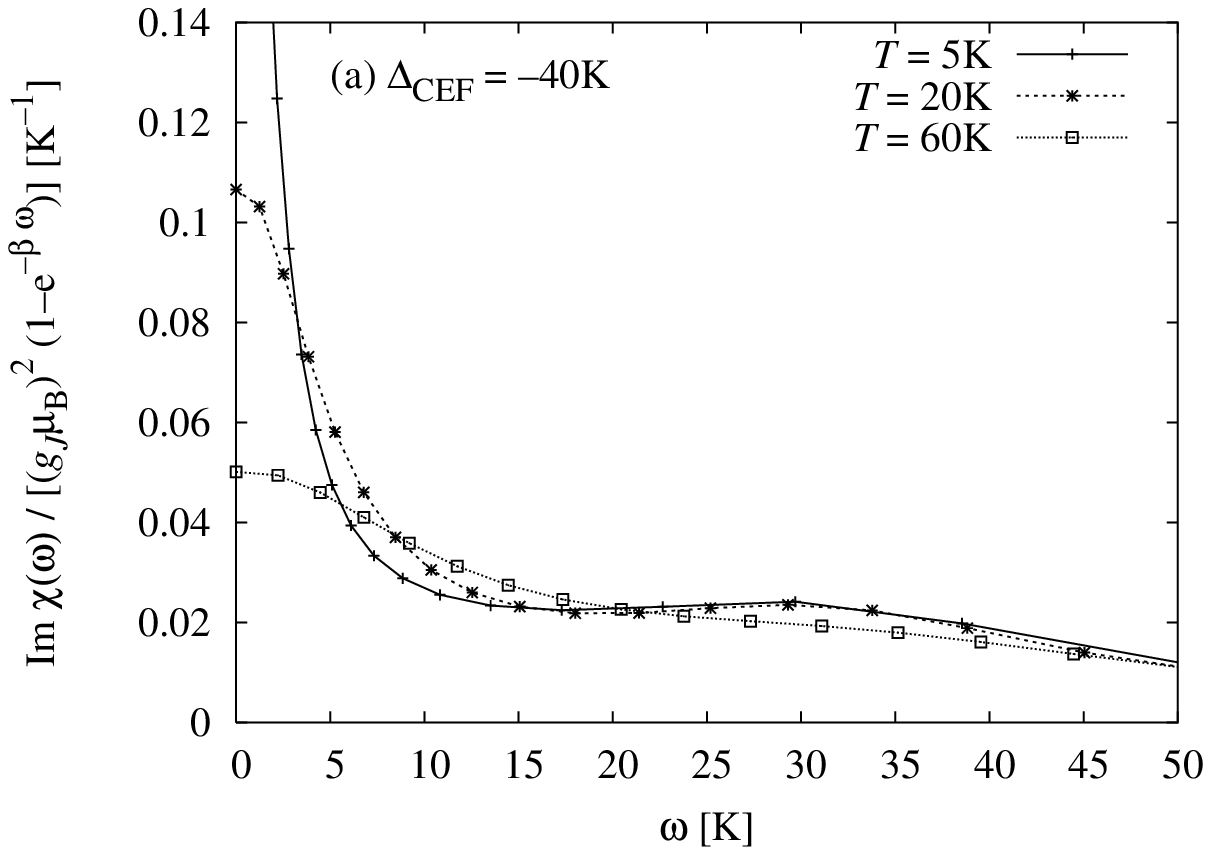}
	\includegraphics[width=7cm]{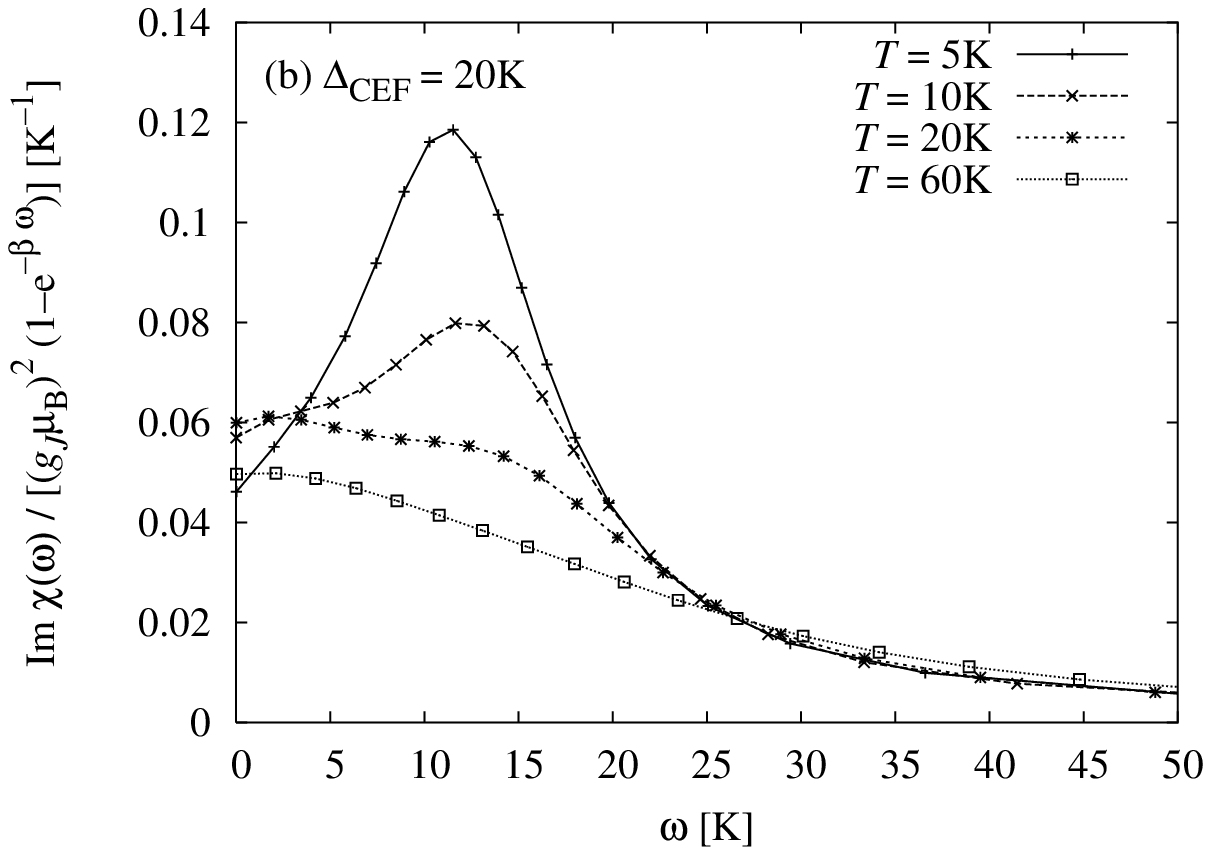}
	\end{center}
	\caption{Temperature dependence of the magnetic spectral intensities. $J_1\rho_c=0.2$, $J_2\rho_c=0$, and CEF splittings $\Delta_{\rm CEF}$ are $-40$K (triplet) in (a) and 20K (singlet) in (b).}
	\label{fig:neutron_CEF}
\end{figure}

We show typical behaviors of magnetic spectra $\text{Im}\chi(\omega)/(1-e^{-\beta \omega})$ both in the CEF singlet and triplet ground states. 
Figure \ref{fig:neutron_CEF} shows the temperature dependence of magnetic spectra. 
In the case of the triplet ground state, quasi-elastic peak develops sharply with decreasing temperature. 
Excitation to the CEF singlet is also found at $\omega\sim 30$K. 
In the case of the CEF singlet ground state, quasi-elastic peak is seen at temperatures higher than $\Delta_{\rm CEF}$, and an inelastic peak develops at lower temperatures. 
The temperature dependence is ascribed to the competition between the CEF singlet and the Kondo effect. Namely, the behavior shows a change from the dominance of the Kondo state at higher temperatures to the CEF singlet at lower temperatures.

\begin{figure}[thb]
	\begin{center}
	\includegraphics[width=7cm]{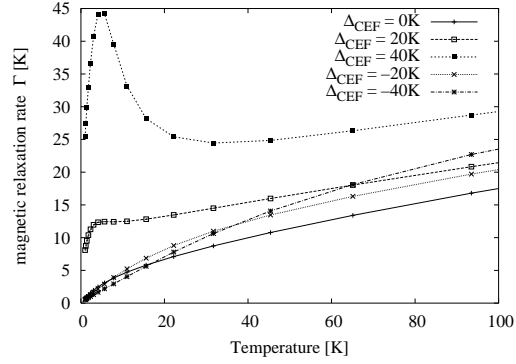}
	\end{center}
	\caption{The magnetic relaxation rate $\Gamma$ as a function of temperature for several values of $\Delta_{\rm CEF}$.}
	\label{fig:relax_rate}
\end{figure}

The magnetic relaxation rate $\Gamma$ measured by NMR is given by
\begin{align}
	\lim_{\omega \rightarrow 0} \frac{\text{Im}\chi(\omega)}{\omega} = \frac{\chi(0)}{\Gamma}.
	\label{eq:relax_rate}
\end{align}
If $\text{Im}\chi(\omega)$ can be approximated by a Lorentzian, $\Gamma$ corresponds to the half-width of a quasi-elastic peak.
Figure \ref{fig:relax_rate} shows the temperature dependence of $\Gamma$ for several values of $\Delta_{\rm CEF}$. 
In the temperature region where the Kondo effect occurs, $\Gamma$ behaves as $T^{a}$ with the exponent $a$ being smaller than unity. 
In the case $\Delta_{\rm CEF}>0$, the crossover to the CEF singlet from the Kondo effect causes an increase of $\Gamma$ as temperature decreases. 
It results from the following reason: $\text{Im}\chi(\omega)/\omega\ |_{\omega \rightarrow 0}$ is reduced at lower temperatures while $\chi(0)$ is constant due to the van Vleck susceptibility in eq. (\ref{eq:relax_rate}). 
The decrease in the case of $\Delta_{\rm CEF}=20$K and 40K below about 5K is caused by the inaccuracy of the NCA at low frequencies, which gives fictitious increase of $\text{Im}\chi(\omega)/\omega$ \cite{nca3}. 
It is known that half-width of $\text{Im}\chi(\omega)/\omega$ at zero temperature corresponds to the Kondo temperature $T_{\rm K}$. We estimate as $T_{\rm K}\sim 1$K for $\Delta_{\rm CEF}=0$ from Fig. \ref{fig:relax_rate}. 

\begin{figure}[thb]
	\begin{center}
	\includegraphics[width=7cm]{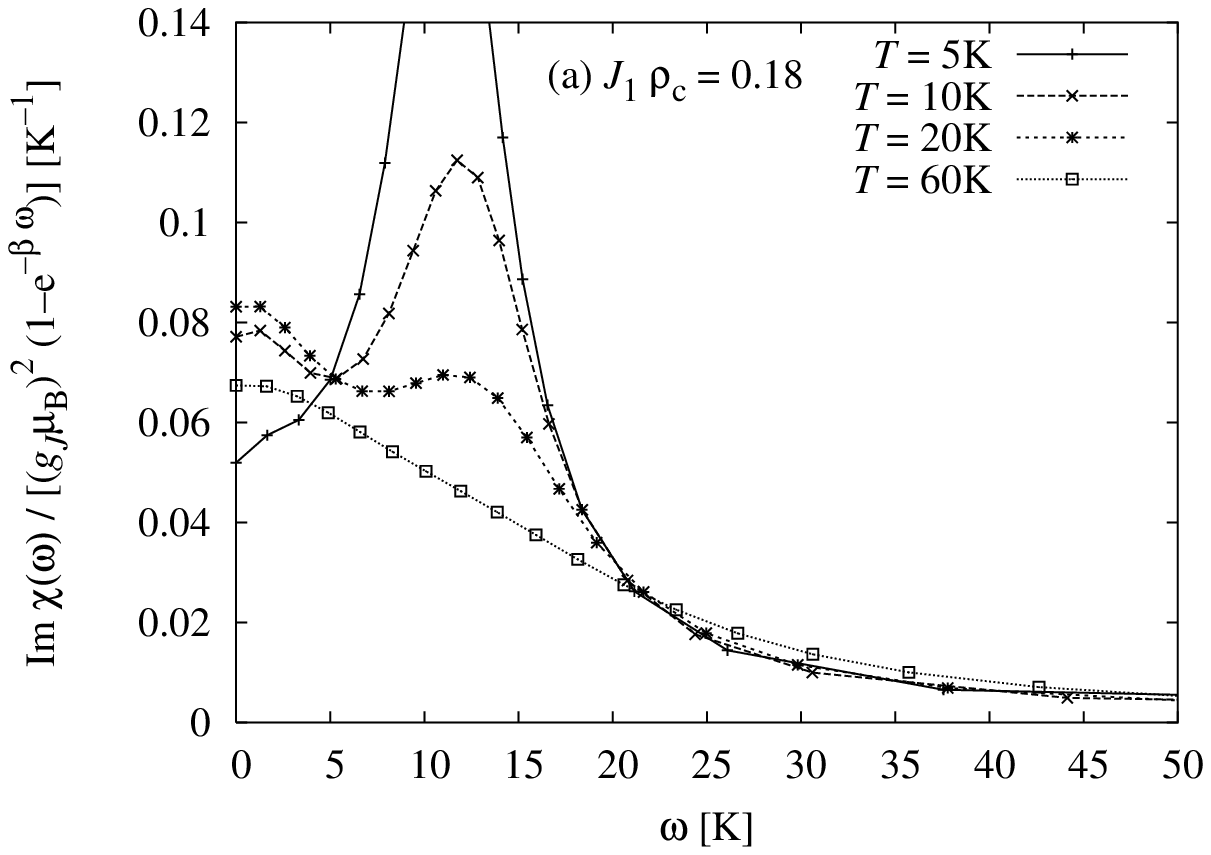}
	\includegraphics[width=7cm]{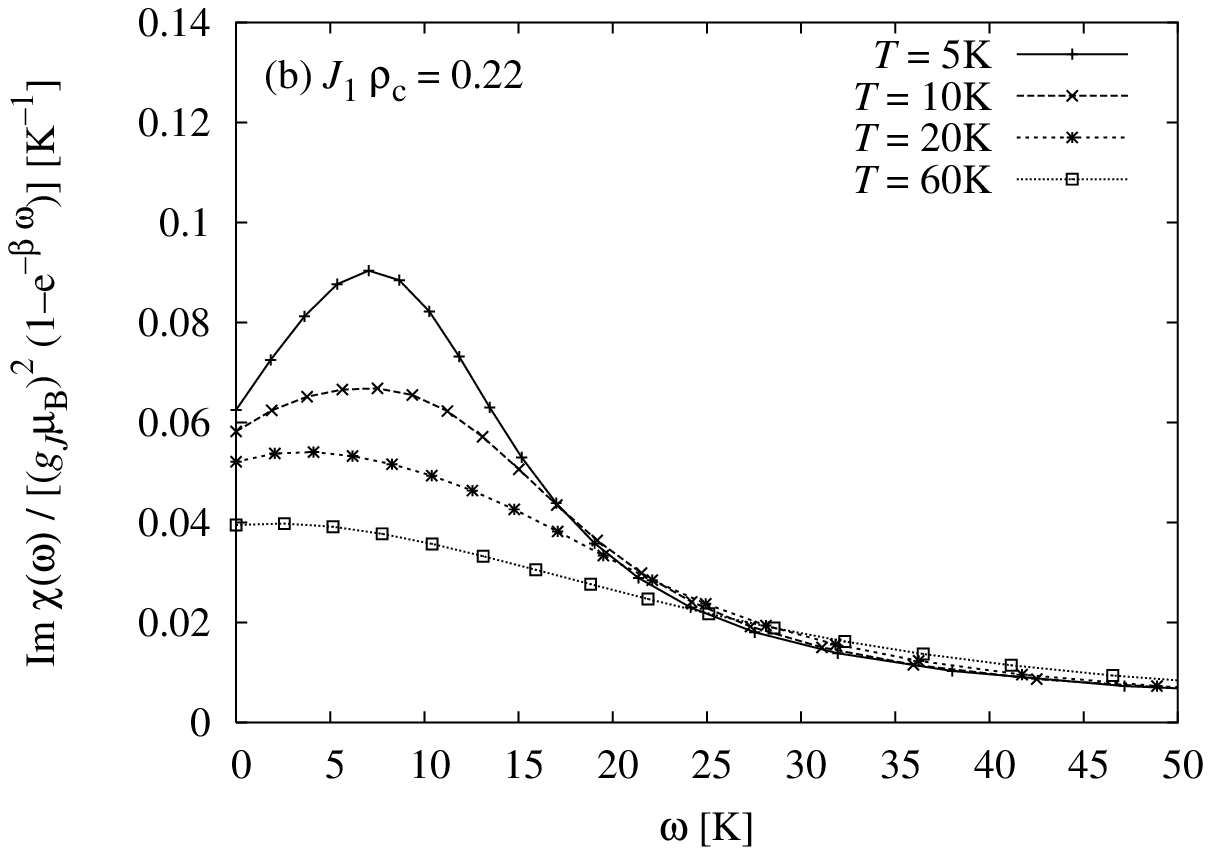}
	\end{center}
	\caption{Temperature dependence of the magnetic spectral intensities with different values for coupling constant $J_1$ but with almost the same value for the renormalized CEF splitting $\tilde{\Delta}_{\rm CEF}\sim10$K. $J_1\rho_c=0.18$, $\Delta_{\rm CEF}=16$K and (b) $J_1\rho_c=0.22$, $\Delta_{\rm CEF}=22$K.}
	\label{fig:neutron_J}
\end{figure}

Next we examine the Kondo temperature $T_{\rm K}$ dependence. Figures \ref{fig:neutron_CEF}(b) and \ref{fig:neutron_J} show magnetic spectra with different values for coupling constant $J_1$ but with renormalized CEF splitting $\tilde{\Delta}_{\rm CEF}$ fixed. 
These parameters satisfy the condition $T_{\rm K}<\tilde{\Delta}_{\rm CEF}$, i.e., no residual entropy. 
In a small coupling case $J_1\rho_c=0.18$, inelastic peak is seen more clearly as compared with the case of $J_1\rho_c=0.20$. 
On the other hand, a large coupling lowers the crossover temperature from the Kondo effect to the CEF singlet. 
Inelastic peak does not arise at all temperatures calculated in Fig. \ref{fig:neutron_J}(b). 
The apparent inelastic peak at $\omega\sim 6$K is actually a quasi-elastic scattering whose maximum is shifted from zero by the Bose factor. 
The crossover temperature depends both on $T_{\rm K}$ and $\tilde{\Delta}_{\rm CEF}$.

\subsection{Application to Pr skutterudites}
We discuss the origin of diversity of the magnetic spectra in PrOs$_4$Sb$_{12}$ and PrFe$_4$P$_{12}$ on the basis of results obtained for the singlet-triplet Kondo model. 
In the case of the CEF level structure for PrOs$_4$Sb$_{12}$, the $4f^2$ configuration has negligible antiferromagnetic exchange interaction with conduction spins because the low lying triplet consists mainly of $\Gamma_5$ in $O_h$\cite{otsuki}. 
We accordingly expect clear CEF excitation, which has actually been observed. 

In the inelastic neutron scattering experiment of PrFe$_4$P$_{12}$, on the other hand, only broad quasi-elastic peak is found at temperatures higher than the phase transition. 
We ascribe the quasi-elastic scattering to the Kondo effect, the occurrence of which is consistent with the temperature dependence of the electrical resistivity\cite{Sato}. 
Inelastic peaks are detected only at lower temperatures. 
The observed inelastic peaks involve a suppression of the Kondo effect. 
The suppression may be due to either the quadrupole order or formation of the CEF singlet. 
We conjecture that the former origin is likely for PrFe$_4$P$_{12}$, in view of recently proposed CEF level scheme, i.e., nearly degenerate singlet-triplet levels\cite{kiss}. 
The CEF splitting in that scheme is not large as compared with the Kondo temperature. Thus the inelastic neutron scattering experiment can resolve the levels only when the quadrupole order breaks the Kondo effect. 
Alternatively the level splitting of the triplet by internal fields is observed. 

\section{Summary}
We have derived dynamics of the CEF singlet-triplet system coupled with conduction spins. 
Explicit calculation is performed with use of the NCA. 
The physical quantities are characterized by two energy scales, i.e., the Kondo temperature $T_{\rm K}$ and renormalized CEF splitting $\tilde{\Delta}_{\rm CEF}$, which is modified from bare one $\Delta_{\rm CEF}$ due to exchange interactions. 

Provided the condition $T_{\rm K}<\tilde{\Delta}_{\rm CEF}$ is satisfied, ground state is the CEF singlet. 
Although $4f^2$ configuration forms the CEF singlet at low temperatures, the Kondo coupling can be present at temperatures higher than $\tilde{\Delta}_{\rm CEF}$. 
Correspondingly, the electrical resistivity is enhanced with decreasing temperature, while it is suppressed by the CEF splitting eventually. 
The competition between the Kondo effect and the CEF singlet also leads to peculiar dependence of the magnetic spectra on temperature. 
As shown in Fig. \ref{fig:neutron_CEF}(b), the inelastic peak cannot be resolved from the broad quasi-elastic peak at $T \gtrsim\tilde{\Delta}_{\rm CEF}$. 
The inelastic peak, however, may be identified as temperature decreases to $T \lesssim\tilde{\Delta}_{\rm CEF}$. 

We have discussed the difference between PrFe$_4$P$_{12}$ and PrOs$_4$Sb$_{12}$ in terms of different wave functions of the CEF triplet states. 
This explains the contrasting behavior observed in the inelastic neutron scattering experiment. 
In PrFe$_4$P$_{12}$, the first excited CEF triplet is composed mainly of $\Gamma_4$ in $O_h$, and antiferromagnetic coupling between the CEF states and conduction spins becomes significant. 
Accordingly, PrFe$_4$P$_{12}$ can exhibit the Kondo effect even with the CEF singlet ground state, and the CEF excitation is suppressed.
Since our calculation does not include intersite interactions, however, we cannot exclude the possibility that the quadrupole order is responsible for suppression of the Kondo effect and the development of the inelastic feature at low temperatures.

%
%
\section*{Acknowledgments}
We acknowledge useful discussions with Prof. K. Iwasa on inelastic neutron scattering results, and with Dr. A. Kiss on the CEF level scheme in PrFe$_4$P$_{12}$. 

%
%
\appendix
\section{Equations for the spectral functions}
Physical quantities are conveniently represented by two types of spectral functions, i.e., $\eta_{\alpha}(\omega)$ and $\xi_{\alpha}(\omega)$. Although $\xi_{\alpha}(\omega)$ is related to $\eta_{\alpha}(\omega)$ analytically, it is difficult to compute directly due to the Boltzmann factor at low temperatures. 
Following ref. \citen{nca3}, we obtain $\xi_{\alpha}(\omega)$ by numerical iteration as follows. 
We introduce an operator $\mathcal{P}^{(\xi)}$ by
$\mathcal{P}^{(\xi)}R(\omega) \equiv -Z_f^{-1}e^{-\beta \omega} \text{Im}R(\omega+i\delta)/ \pi$. 
Operating $\mathcal{P}^{(\xi)}$ on eqs. (\ref{eq:self}) and (\ref{eq:pi}), we obtain
\begin{align}
	\sigma_{\alpha}(\omega) &\equiv \mathcal{P}^{(\xi)} \Sigma_{\alpha}(\omega)
	 = -2 \int d\epsilon \rho_c(\epsilon) f(\epsilon) \xi_{\alpha}^{(K)} (\omega-\epsilon), \\
	\pi_{\alpha}(\omega) &\equiv \mathcal{P}^{(\xi)}\Pi_{\alpha}(\omega)
	 = \int d \epsilon \rho_c (\epsilon) [1-f(\epsilon)] \xi_{\alpha}(\omega+\epsilon). 
\end{align}
Here $\xi_{\alpha}(\omega)$ are related with the corresponding resolvent as
\begin{align}
	\xi_{\alpha}(\omega) &\equiv \mathcal{P}^{(\xi)} R_{\alpha}(\omega)
	 = |R_{\alpha}(\omega+i\delta)|^2 \sigma_{\alpha} (\omega).
\end{align}
Analogously, we obtain 
\begin{align}
	\xi_{\text{t}}^{(I)} &\equiv \mathcal{P}^{(\xi)}\tilde{I}_{\rm t}
	 = \frac{|\tilde{I}_{\rm t} |^2}{|c_1 |^2}
	 [ (\mathcal{P}^{(\xi)}a^*) b^* c_1 + a(\mathcal{P}^{(\xi)}b^*) c_1 + ab (\mathcal{P}^{(\xi)}c_1) ], \nonumber \\
	\xi_{\text{s}}^{(I)} &\equiv \mathcal{P}^{(\xi)}\tilde{I}_{\rm s}
	 = \frac{|\tilde{I}_{\rm s} |^2}{c_2} (\mathcal{P}^{(\xi)}b^*), \nonumber \\
	\xi_{\text{t}}^{(K)} &\equiv \mathcal{P}^{(\xi)}\tilde{K}_{\rm t}
	 = -\frac{|\tilde{K}_{\text{t}} |^2}{|c_3 |^2}
	 [ (\mathcal{P}^{(\xi)}a^*) b^* c_3 + a(\mathcal{P}^{(\xi)}b^*) c_3
	  + ab (\mathcal{P}^{(\xi)}c_3) ], \nonumber \\
	\xi_{\text{s}}^{(K)} &\equiv \mathcal{P}^{(\xi)}\tilde{K}_{\rm s}
	 = -\frac{|\tilde{K}_{\text{s}} |^2}{|c_4 |^2}
	 [ (\mathcal{P}^{(\xi)}b^*) c_4 + b (\mathcal{P}^{(\xi)}c_4) ],
\end{align}
where we have used the notations in eq. (\ref{eq:renorm_interaction}) and
\begin{align}
	\mathcal{P}^{(\xi)}a^* &= I_{\rm t} \pi_{\text{t}}, \nonumber \\
	\mathcal{P}^{(\xi)}b^* &= -4I_{\rm t} \pi_{\text{t}}
	 + 3 I_{\rm s}^2 (\pi_{\text{t}}\Pi_{\text{s}}+\Pi_{\text{t}}\pi_{\text{s}}), \nonumber \\
	\mathcal{P}^{(\xi)}c_1 &= -4I_{\rm s}^2\pi_{\text{s}}, \nonumber \\
	\mathcal{P}^{(\xi)}c_3 &= 4I_{\rm t}^2 \pi_{\text{t}} + 2I_{\rm s}^2 \pi_{\text{s}}
	 -3 I_{\rm t} I_{\rm s}^2 (\pi_{\text{t}}\Pi_{\text{s}}+\Pi_{\text{t}}\pi_{\text{s}}), \nonumber \\
	\mathcal{P}^{(\xi)}c_4 &= 3 I_{\rm s}^2 \pi_{\text{t}}.
\end{align}
The norms are determined by the following sum rule: 
\begin{align}
	\sum_{\alpha} \int_{-\infty}^{\infty} d\omega \xi_{\alpha}(\omega) = 1.
	\label{eq:sum_rule}
\end{align}
Normalization by eq. (\ref{eq:sum_rule}) is equivalent to determination of the partition function $Z_f$ of $f$ electrons.

\section{Equations for the magnetic susceptibility}
The formulae of $\text{Im}\chi(\omega)$ in eq. (\ref{eq:Im_chi}) and $\chi(0)$ in eq. (\ref{eq:chi0}) include integrations with the Boltzmann factor. To avoid this factor, we rewrite the equations following ref. \citen{nca4}. 
For each frequency $\omega$ we introduce the following simplifying notations:
\begin{align}
	Q_{\lambda}(\epsilon) &= \Lambda_{\lambda} (\epsilon + i\delta, \epsilon + \omega+i\delta), \nonumber \\
	\tilde{Q}_{\lambda}(\epsilon) &= \Lambda_{\lambda} (\epsilon-i\delta, \epsilon+\omega+i\delta), \nonumber \\
	R_{\alpha \alpha'}(\epsilon) &= R_{\alpha} (\epsilon + i\delta)
	 R_{\alpha'}(\epsilon + \omega+i\delta), \nonumber \\
	\tilde{R}_{\alpha \alpha'}(\epsilon) &= R_{\alpha} (\epsilon -i\delta)
	 R_{\alpha'}(\epsilon + \omega+i\delta). 
\end{align}
$I_{\alpha \alpha'}(\epsilon)$, $\tilde{I}_{\alpha \alpha'}(\epsilon)$, $K_{\alpha \alpha'}(\epsilon)$ and $\tilde{K}_{\alpha \alpha'}(\epsilon)$ are defined with use of $\tilde{I}_{\alpha}(z)$ or $\tilde{K}_{\alpha}(z)$ in a similar manner. 
We introduce an operator $\mathcal{P}$ by $\mathcal{P}Q(\epsilon) \equiv Z_f^{-1} e^{-\beta \epsilon} [\tilde{Q}(\epsilon)-Q(\epsilon)]$. 
In terms of $\mathcal{P}$, eq. (\ref{eq:Im_chi}) is represented by
\begin{align}
	\text{Im} \chi_{\lambda}(\omega +i\delta)
	 =  (1- e^{-\beta \omega}) \int_{-\infty}^{\infty} \frac{d\epsilon}{2\pi}
	  \text{Re} \mathcal{P} Q_{\lambda}(\epsilon),
\end{align}
without the Boltzmann factor. 
Operating $\mathcal{P}$ on eqs. (\ref{eq:ver_t}), (\ref{eq:ver_sp}) and (\ref{eq:ver_sc}), 
we obtain
\begin{align}
	&\mathcal{P}Q_{\rm t}(\epsilon) = \mathcal{P}R_{\rm tt}(\epsilon) \nonumber \\
	 &+ [\mathcal{P}R_{\rm tt}(\epsilon)] 2\int d\epsilon' \rho_c(\epsilon') [1-f(\epsilon')]
	 \int d\epsilon'' \rho_c(\epsilon'') f(\epsilon'')
	 [\frac{1}{4} I_{\rm tt} (\epsilon-\epsilon')+ K_{\rm tt} (\epsilon-\epsilon')] Q_{\rm t}(\epsilon-\epsilon'+\epsilon'') \nonumber \\
	 &+ \tilde{R}_{\rm tt}(\epsilon) 2\int d\epsilon' \rho_c(\epsilon') f(\epsilon')
	 \int d\epsilon'' \rho_c(\epsilon'') f(\epsilon'')
	 [\frac{1}{4} \mathcal{P}I_{\rm tt} (\epsilon-\epsilon')+ \mathcal{P}K_{\rm tt} (\epsilon-\epsilon')]
	 Q_{\rm t}(\epsilon-\epsilon'+\epsilon'') \nonumber \\
	 &+ \tilde{R}_{\rm tt}(\epsilon) 2\int d\epsilon' \rho_c(\epsilon') f(\epsilon')
	 \int d\epsilon'' \rho_c(\epsilon'') [1-f(\epsilon'')]
	 [\frac{1}{4} \tilde{I}_{\rm tt} (\epsilon-\epsilon')+ \tilde{K}_{\rm tt} (\epsilon-\epsilon')] [\mathcal{P}Q_{\rm t}(\epsilon-\epsilon'+\epsilon'')],
	 \label{eq:pq_t}
\end{align}
\begin{align}
	&\mathcal{P}Q_{\rm sp}(\epsilon) = \mathcal{P}R_{\rm st}(\epsilon) \nonumber \\
	 &+ [\mathcal{P}R_{\rm st}(\epsilon)] 2\int d\epsilon' \rho_c(\epsilon') [1-f(\epsilon')]
	 \int d\epsilon'' \rho_c(\epsilon'') f(\epsilon'')
	 K_{\rm st} (\epsilon-\epsilon') Q_{\rm sp}(\epsilon-\epsilon'+\epsilon'') \nonumber \\
	 &+ \tilde{R}_{\rm st}(\epsilon) 2\int d\epsilon' \rho_c(\epsilon') f(\epsilon')
	 \int d\epsilon'' \rho_c(\epsilon'') f(\epsilon'')
	 [\mathcal{P}K_{\rm st} (\epsilon-\epsilon')]
	 Q_{\rm sp}(\epsilon-\epsilon'+\epsilon'') \nonumber \\
	 &+ \tilde{R}_{\rm st}(\epsilon) 2\int d\epsilon' \rho_c(\epsilon') f(\epsilon')
	 \int d\epsilon'' \rho_c(\epsilon'') [1-f(\epsilon'')]
	 \tilde{K}_{\rm st} (\epsilon-\epsilon') [\mathcal{P}Q_{\rm sp}(\epsilon-\epsilon'+\epsilon'')] \nonumber \\
	 &+ [\mathcal{P}R_{\rm st}(\epsilon)] 2\int d\epsilon' \rho_c(\epsilon') [1-f(\epsilon')]
	 \int d\epsilon'' \rho_c(\epsilon'') f(\epsilon'')
	 I_{\rm ss} (\epsilon-\epsilon') Q_{\rm sc}(\epsilon-\epsilon'+\epsilon'') \nonumber \\
	 &+ \tilde{R}_{\rm st}(\epsilon) 2\int d\epsilon' \rho_c(\epsilon') f(\epsilon')
	 \int d\epsilon'' \rho_c(\epsilon'') f(\epsilon'')
	 [\mathcal{P}I_{\rm ss} (\epsilon-\epsilon')]
	 Q_{\rm sc}(\epsilon-\epsilon'+\epsilon'') \nonumber \\
	 &+ \tilde{R}_{\rm st}(\epsilon) 2\int d\epsilon' \rho_c(\epsilon') f(\epsilon')
	 \int d\epsilon'' \rho_c(\epsilon'') [1-f(\epsilon'')]
	 \tilde{I}_{\rm ss} (\epsilon-\epsilon') [\mathcal{P}Q_{\rm sc}(\epsilon-\epsilon'+\epsilon'')],
	 \label{eq:pq_sp}
\end{align}
\begin{align}
	&\mathcal{P}Q_{\rm sc}(\epsilon) \nonumber \\
	 &= [\mathcal{P}R_{\rm ts}(\epsilon)] 2\int d\epsilon' \rho_c(\epsilon') [1-f(\epsilon')]
	 \int d\epsilon'' \rho_c(\epsilon'') f(\epsilon'')
	 K_{\rm ts} (\epsilon-\epsilon') Q_{\rm sc}(\epsilon-\epsilon'+\epsilon'') \nonumber \\
	 &+ \tilde{R}_{\rm ts}(\epsilon) 2\int d\epsilon' \rho_c(\epsilon') f(\epsilon')
	 \int d\epsilon'' \rho_c(\epsilon'') f(\epsilon'')
	 [\mathcal{P}K_{\rm ts} (\epsilon-\epsilon')]
	 Q_{\rm sc}(\epsilon-\epsilon'+\epsilon'') \nonumber \\
	 &+ \tilde{R}_{\rm ts}(\epsilon) 2\int d\epsilon' \rho_c(\epsilon') f(\epsilon')
	 \int d\epsilon'' \rho_c(\epsilon'') [1-f(\epsilon'')]
	 \tilde{K}_{\rm ts} (\epsilon-\epsilon') [\mathcal{P}Q_{\rm sc}(\epsilon-\epsilon'+\epsilon'')] \nonumber \\
	 &+ [\mathcal{P}R_{\rm ts}(\epsilon)] 2\int d\epsilon' \rho_c(\epsilon') [1-f(\epsilon')]
	 \int d\epsilon'' \rho_c(\epsilon'') f(\epsilon'')
	 I_{\rm ss} (\epsilon-\epsilon') Q_{\rm sp}(\epsilon-\epsilon'+\epsilon'') \nonumber \\
	 &+ \tilde{R}_{\rm ts}(\epsilon) 2\int d\epsilon' \rho_c(\epsilon') f(\epsilon')
	 \int d\epsilon'' \rho_c(\epsilon'') f(\epsilon'')
	 [\mathcal{P}I_{\rm ss} (\epsilon-\epsilon')]
	 Q_{\rm sp}(\epsilon-\epsilon'+\epsilon'') \nonumber \\
	 &+ \tilde{R}_{\rm ts}(\epsilon) 2\int d\epsilon' \rho_c(\epsilon') f(\epsilon')
	 \int d\epsilon'' \rho_c(\epsilon'') [1-f(\epsilon'')]
	 \tilde{I}_{\rm ss} (\epsilon-\epsilon') [\mathcal{P}Q_{\rm sp}(\epsilon-\epsilon'+\epsilon'')].
	 \label{eq:pq_sc}
\end{align}
$\mathcal{P}R_{\alpha \alpha'}(\epsilon)$ is given by known quantities as follows:
\begin{align}
	\mathcal{P}R_{\alpha \alpha'}(\epsilon) &= 2\pi i \xi_{\alpha}(\epsilon)
	 R_{\alpha'}(\epsilon + \omega+i\delta).
\end{align}
$\mathcal{P}I_{\alpha \alpha'}(\epsilon)$ and $\mathcal{P}K_{\alpha \alpha'}(\epsilon)$ are obtained by analogous equations. 
After $Q_{\gamma}$ are obtained using eqs. (\ref{eq:ver_t}), (\ref{eq:ver_sp}) and (\ref{eq:ver_sc}), $\mathcal{P}Q_{\gamma}$ are computed by numerical iterations of eqs. (\ref{eq:pq_t}), (\ref{eq:pq_sp}) and (\ref{eq:pq_sc}).

For the static susceptibility, we introduce another operator $\hat{P}$ by $\hat{P}Q(\epsilon)\equiv Z_f^{-1}e^{-\beta \epsilon}\text{Im}Q(\epsilon)$ instead of $\mathcal{P}$. 
The static susceptibility is given in terms of $\hat{\mathcal{P}}$ as follows:
\begin{align}
	\chi_{\gamma}(0) = \int_{-\infty}^{\infty} \frac{d\epsilon}{\pi}
	 \hat{\mathcal{P}}Q_{\gamma}(\epsilon).
\end{align}
The equations for $\hat{\mathcal{P}}Q_{\gamma}$ are obtained by replacing 
$\mathcal{P}, \tilde{R}_{\alpha \alpha'}, \tilde{I}_{\alpha \alpha'}, \tilde{K}_{\alpha \alpha'}$ by $\hat{\mathcal{P}}, R^*_{\alpha \alpha'}, I^*_{\alpha \alpha'}, K^*_{\alpha \alpha'}$
in eqs. (\ref{eq:pq_t}), (\ref{eq:pq_sp}) and (\ref{eq:pq_sc}). 
We also obtain $\hat{\mathcal{P}} R_{\alpha \alpha'}(\epsilon)$ by
\begin{align}
	\hat{\mathcal{P}} R_{\alpha \alpha'}(\epsilon)
	&= -\pi [\xi_{\alpha}(\epsilon) R_{\alpha'}(\epsilon+i\delta)
	+ R_{\alpha}^{*}(\epsilon+i\delta) \xi_{\alpha'}(\epsilon)]. 
	\label{eq:hat_P_R}
\end{align}
The formulae for $\hat{\mathcal{P}} I_{\alpha \alpha'}(\epsilon)$ and $\hat{\mathcal{P}} K_{\alpha \alpha'}(\epsilon)$ are obtained by analogy of eq. (\ref{eq:hat_P_R}). 
$\hat{\mathcal{P}}Q_{\gamma}$ are numerically computed in a way similar to $\mathcal{P}Q_{\gamma}$.

\end{document}